\documentstyle[preprint,aps]{revtex}
\begin{document}
\draft
\renewcommand{\baselinestretch}{1.2}
\preprint{}
\title{Role of the $a_1$ meson in dilepton production
\\from hot hadronic matter}
\author{Chungsik Song and Che Ming Ko}
\address{ Cyclotron Institute,
        Texas A\&M University, College Station, TX 77843 \\}
\author{Charles Gale}
\address{ Physics Department, McGill University\\
         3600 University St., Montr\'eal, QC, H3A 2T8, Canada\\}
\date{\today}
\maketitle
\begin{abstract}
Dilepton production from hot hadronic matter is studied in an
effective chiral Lagrangian with pions, $\rho$-mesons, and $a_1$ mesons,
We find that the production rates from reactions that involve axial-vector
mesons dominate over contributions from all other reactions
when the dilepton  invariant mass is above 1.5 GeV.
\end{abstract}


In  nucleus-nucleus collisions at ultrarelativistic energies, a hot
and dense matter consisting of quarks and gluons
is expected to be formed in the initial
stages of the collision.  This quark-gluon plasma will, however,
transform into hadronic matter as it expands and cools below the
critical temperature.  Since dileptons and photons produced from
the quark-gluon plasma do not suffer final-state interactions, they carry
information about their production  and have been considered as possible
signatures for the formation of
the quark-gluon plasma \cite{fei,shur78,larray,drell}.

However, hadrons in the hadronic matter
still interact violently until freeze out.
Dileptons and photons can thus be produced also from
the hot and dense hadronic matter.   To use dileptons and photons
as signatures for the quark-gluon plasma, we need therefore
to distinguish them from those produced from the hadronic matter.

For dileptons with invariant masses larger than the $J/\psi$
mass (3 GeV), the dominant
contributions are from the Drell-Yan process and
direct charm decay \cite{drell}, while for
invariant masses lower than the phi meson mass (1 GeV),
radiative and direct decays, together with
$\pi\pi$ annihilations, form the brightest source.
To distinguish the quark-gluon plasma with these  low invariant mass lepton
pairs is thus expected to be extremely
difficult \cite{kevin}.  However,
for dileptons of invariant masses that are between the
phi and $J/\Psi$ masses, $m_\phi<M<m_{J/\Psi}$, the original suggestion was
that
the contribution from
the quark-gluon plasma may dominate that from the hadronic matter
\cite{shur78}. One may then be able to study the quark-gluon plasma
by concentrating on those lepton pairs.

The dilepton production rates from hadronic matter
have usually been calculated by assuming that the
hadronic matter consists of only pions.
Recently, one of us and P.~Lichard
\cite{gale} have
shown that for temperatures $T > 100$ MeV dilepton production from
reactions involving higher-mass hadron resonances becomes important.  Including
both
strange and non-strange pseudoscalar and vector mesons,
it was found that in the invariant mass region where one could expect
dominant dilepton contributions from the quark-gluon plasma,
the reactions involving pseudoscalar and vector mesons lead to
significant lepton pair signals.
It is thus essential to carefully study dilepton production
from hot hadronic matter with the inclusion of higher mass mesons.

In connection with photon production from a hot hadronic gas, it has recently
been shown that the contributions from processes with $a_1$ mesons
in the intermediate states dominate \cite{xsb,song1}.
It is thus reasonable to expect that the $a_1$ meson would also play an
important role
in dilepton production. It is the purpose of this paper to investigate this
assertion. We shall study dilepton production from
processes involving the $a_1$ meson in thermalized hadronic matter using
an effective chiral
Lagrangian that includes not only pseudoscalar and vector mesons
but also axial-vector mesons \cite{song1,song2,song3}.

In our effective Lagrangian, the
pseudoscalar mesons ($\phi$) are described by the non-linear
$\sigma$ model while the vector ($V_\mu$) and axial-vector ($A_\mu$) mesons
are included as massive Yang-Mills fields of the $SU(2)\times SU(2)$
chiral symmetry.
The lowest order interaction terms are given by
\begin{eqnarray}
{\cal L}^{(3)}_{V\phi\phi}&=
&{ig\over 2} {\rm Tr}{\partial_\mu \phi[V^\mu,\phi]}\cr
&&+{ig \delta\over 2m_V^2} {\rm Tr}{(\partial_\mu V_\nu
                    -\partial_\nu V_\mu) \partial^\mu\phi \partial^\nu
\phi},\cr
{\cal L}^{(3)}_{VVV}&=
 & {ig\over 2}{\rm Tr}{(\partial_\mu V_\nu-\partial_\nu V_\mu) V^\mu V^\nu},\cr
{\cal L}^{(3)}_{VA\phi}&=
&i\left[\left({1-\sigma\over1+\sigma}\right)^{1/2}{g^2 F_\pi\over 4m_V^2}
+{2\xi g Z^2\over F_\pi\sqrt{1+\sigma}}\right]
{\rm Tr}(\partial_\mu V_\nu-\partial_\nu V_\mu)[A^\mu,\partial^\nu\phi]\cr
&&+i\left[\left({1+\sigma \over1-\sigma}\right)^{1/2}{g^2 F_\pi\over 4m_V^2}
-{2\sigma\over\ F_\pi\sqrt{1-\sigma^2}}\right]
{\rm Tr}(\partial_\mu A_\nu -\partial_\nu A_\mu)[\partial^\mu V^\nu, \phi],\cr
{\cal L}^{(3)}_{VAA}&=
&{ig\over 2}{\rm Tr}{(\partial_\mu A_\nu-\partial_\nu A_\mu)
                          [V^\mu,A^\nu]}\cr
&&+{ig\over 2}\left({1-\sigma\over 1+\sigma}\right)
                 \left(1-{2\xi g\over \sqrt{1-\sigma}}\right)
                 {\rm Tr}{(\partial_\mu V_\nu-\partial_\nu V_\mu)A^\mu A^\nu},
\end{eqnarray}
where $F_\pi\approx 135$ MeV is the pion decay constant,
\begin{equation}
\delta={g^2F_\pi^2\over 4m_V^2}-{2\xi g\over\sqrt{1-\sigma}}
       {4m_V^2\over g^2F_\pi^2}Z^4,
\end{equation}
and
\begin{equation}
Z^2=1-\left({g^2F_\pi^2\over 4 m_V^2}\right).
\end{equation}
The parameters of the effective Lagrangian ($g, \xi, \sigma, m_V$) have been
determined from the experimental data on the decay widths and masses of
$\rho$ and $a_1$ mesons.

The electromagnetic interaction is introduced through
imposing the $U(1)_{EM}$ gauge symmetry on the effective chiral
Lagrangian, i.e., we let
\begin{equation}
{\cal L}\to {\cal L}-{1\over 4}(\partial_\mu a_\nu-\partial_\nu a_\mu)^2
+{\cal L}_{EM},
\end{equation}
where $a_\mu$ is the electromagnetic field and
\begin{eqnarray}
{\cal L}_{EM}&=&-{2e m_V^2\over g} a_\mu {\rm Tr}[Q V_\mu]
+{2 e^2 m_V^2\over g^2}a_\mu^2 {\rm Tr}Q^2\cr
&=&-{\sqrt{2}e\over g}a^\mu\left[m_\rho^2\rho^0_\mu
+{1\over3}m_\omega^2\omega_\mu-{\sqrt{2}\over3}m_\phi^2\phi_\mu\right]
+{\cal O}(a_\mu^2),
\end{eqnarray}
with $Q$=dial(${2\over3}$,$-{1\over3}$,$-{1\over3}$) is the quark charge
matrix.  ${\cal L}_{EM}$
ensures that  the Lagrangian is  invariant under  the gauge transformation.
The above equation shows that the electromagnetic couplings of the
vector mesons are given by the same form as in the vector meson dominance
model \cite{vmd}.


The $a_1$ meson will contribute to dilepton production mainly through
the two processes,
$\pi^++a_1^-(\pi^-+a_1^+)\to e^++e^-$ and $a_1^++a_1^-\to e^++e^-$,
as shown by the two diagrams in Fig.~1.
While the annihilation of two $a_1$ mesons
yields a lepton pair with a large invariant mass,
the annihilation of the $a_1$ meson with a pion
will be of interest in the invariant mass region considered here.

In general, the dilepton production rate from the annihilation of
two hadrons $h_1$ and $h_2$ can be written as
\begin{equation}
{dN\over d^4x}={\cal N}\int{d^3p_1\over(2\pi)^3}{d^3p_2\over(2\pi)^3}
               f(p_1)f(p_2)\sigma(1+2\to l^+l^-)v_{\rm rel},
\end{equation}
where ${\cal N}$ is an overall degeneracy factor, $f(p)$ is the distribution
function of the incoming particles at temperature $T$, $v_{\rm rel}$ is the
relative velocity of the two particles, and $\sigma$ is the
dilepton production cross section for
the reaction $h_1+h_2 \to l^++l^-$ \cite{larray}.
As the cross section depends only on the square of the invariant mass
$s=(p_1+p_2)^2=M^2$, this expression can be simplified to
\begin{equation}
{dN\over d^4x}={\cal N}{T^2\over2(2\pi)^4}\int_{s_0}ds\sigma(s)
\sqrt{s^2-2s(m_1^2+m_2^2) +(m_1^2-m_2^2)^2}\,G(T,s),
\end{equation}
where
\begin{equation}
G(T,s)=\int_{m_1/T}dx{1\over e^x-1}\ln\left({1-\exp(-y_+)\over 1-\exp(-y_-)}
\right),
\end{equation}
with
\begin{equation}
y_\pm={1\over 2m_1^2}
\left[(s-m_1^2-m_2^2)x\pm\sqrt{s^2-2s(m_1^2+m_2^2)+(m_1^2-m_2^2)^2}
\sqrt{x^2-{m_1^2\over T^2}}\,\right].
\end{equation}
In this calculation, we focus on dielectrons and have neglected the small
lepton masses but it is
straightforward to include them.  The differential production
rate then becomes
\begin{equation}
{dN\over d^4xdM^2}={\cal N}{T^2\over2(2\pi)^4}\sigma(M^2)
\sqrt{M^4-2M^2(m_1^2+m_2^2) +(m_1^2-m_2^2)^2}\,G(T,M^2),
\end{equation}
where $M$ is the invariant mass of the dilepton.

The dilepton production cross section is given by the product of a form
 factor and the square of a scattering amplitude, which can be written as
\begin{equation}
\bar{\vert{\cal M}\vert^2}=4\left({4\pi\alpha\over q^2}\right)^2
L_{\mu\nu}H^{\mu\nu},
\end{equation}
with $q=p_1+p_2=p_3+p_4$ and $\alpha$ the fine structure constant.
In the above, $L_{\mu\nu}$ is the leptonic tensor given by
\begin{equation}
L^{\mu\nu}=p_3^\mu p_4^\nu+p_4^\mu p_3^\nu-g^{\mu\nu}p_3\cdot p_4,
\end{equation}
and $H^{\mu\nu}$ is a hadronic tensor for the reaction.


The hadronic tensor $H^{\mu\nu}$ for the reaction $\pi^++\pi^-\to e^++e^-$
is given by
\begin{equation}
H^{\mu\nu}=(p_2^\mu-p_1^\mu)(p_2^\nu-p_1^\nu),
\end{equation}
which leads to the well-known result for the $\pi \pi$ annihilation cross
section
\begin{equation}
\sigma_\pi(s)={4\pi\over3}{\alpha^2\over s}\vert F_\pi\vert^2\left(1-{4
m_\pi^2\over s}\right)^{1/2},
\end{equation}
where $F_\pi$ is the pion form factor.


For the reaction $\rho^+\rho^-\to e^+e^-$, the hadronic tensor
$H^{\mu\nu}$ is
\begin{eqnarray}
H^{\mu\nu}&=&h_\rho^{\mu\alpha\beta} h_{\rho\,\alpha\beta}^\nu
-h_\rho^{\mu\alpha\beta}p_{1\,\beta}
 h_{\rho\,\alpha}^{\nu\beta}p_{1\,\beta}/m_\rho^2\cr
&&-h_\rho^{\mu\alpha\beta}p_{2\,\alpha}
 h_{\rho\,\beta}^{\nu\alpha}p_{2\,\alpha}/m_\rho^2
+h_\rho^{\mu\alpha\beta}p_{1\,\beta}p_{2\,\alpha}
 h_\rho^{\mu\alpha\beta}p_{2\,\alpha}p_{1\,\beta}/(m_\rho^2)^2,
\end{eqnarray}
with
\begin{equation}
h_\rho^{\mu\alpha\beta}=(p_2^\mu-p_1^\mu)g^{\alpha\beta}
                  +(q^\alpha-p_2^\alpha)g^{\beta\mu}
                  +(p_1^\beta-q^\beta)g^{\mu\alpha}.
\end{equation}


For $\pi a_1$ annihilation into a lepton pair,
the hadronic tensor has the form
\begin{equation}
H^{\mu\nu}=h_{a_1}^{\mu\alpha} h_{a_1\,\alpha}^\nu
-h_{a_1}^{\mu\alpha}p_{2\,\alpha}
 h_{a_1}^{\nu\alpha}p_{2\,\alpha}/m_{a_1}^2,
\end{equation}
with
\begin{equation}
h_{a_1}^{\mu\alpha}=\eta_1[(p_1\cdot q)g^{\mu\alpha}-p_1^\mu q^\alpha]
                   +\eta_2[(p_2\cdot q)g^{\mu\alpha}-p_2^\mu q^\alpha],
\end{equation}
where
\begin{eqnarray}
\eta_1&=&\left({1-\sigma\over 1+\sigma}\right)^{1/2}\left(gF_\pi\over 2m_\rho^2
\right)+{4\xi Z^2\over F_\pi\sqrt{1+\sigma}},\cr
\eta_2&=&\left({1+\sigma\over 1-\sigma}\right)^{1/2}\left(gF_\pi\over 2m_\rho^2
\right)-{4\sigma\over gF_\pi\sqrt{1-\sigma^2}}.
\end{eqnarray}


The hadronic tensor for the reaction  $a_1^++a_1^-\to e^++e^-$
can be obtained from the reaction $\rho^+\rho^-\to e^+e^-$
by replacing $h_\rho$ and $m_\rho$ with $h_{a_1}$ and $m_{a_1}$,
respectively. The $h_{a_1}^{\mu\alpha\beta}$ is then given by
\begin{equation}
h_{a_1}^{\mu\alpha\beta}=(p_2^\mu-p_1^\mu)g^{\alpha\beta}
                  +(\zeta q^\alpha-p_2^\alpha)g^{\beta\mu}
                  +(p_1^\beta-\zeta q^\beta)g^{\mu\alpha},
\end{equation}
with
\begin{equation}
\zeta=\left({1-\sigma\over 1+\sigma}\right)
      \left(1-{2\xi g\over\sqrt{1-\sigma}}\right).
\end{equation}


We have calculated the dilepton production rate from hadronic matter
at two temperatures, $T$=150 and 200 MeV,  and
the results are shown in Fig.~2 and Fig.~3, respectively.
The dotted curve is the ``usual'' result from
the reaction $\pi^+\pi^-\to e^+e^-$
which dominates at low invariant masses.
The contribution from the reaction  $\rho^+\rho^-\to e^+e^-$
is shown by the dashed curve. This process was found in ref. \cite{gale} to
be the dominant one for
dileptons of invariant masses in the region 1.5 GeV $< M <$ 3.0 GeV.
The solid curves are from the reactions
$\pi^+a_1^-(\pi^-a_1^+)\to e^+e^-$ and $a_1^+a_1^-\to e^+e^-$,
both involving the $a_1$ meson.
We note that in our calculations
we have used the same form factors for $\pi^+\pi^-\to e^+e^-$
and $\rho^+\rho^-\to e^+e^-$, as in ref. \cite{gale}.
For the reactions $\pi^+a_1^-(\pi^-a_1^+)\to e^+e^-$ and
$a_1^+a_1^-\to e^+e^-$,
we assume that the form factors are the same as for $\pi^+\pi^-$
annihilation. This is again consistent with the prescription followed in  ref.
\cite{gale}.

Our results at $T=150$ MeV show that the $a_1$ meson plays an important role
in dilepton production  from hadronic matter.
Dilepton production from  $\pi^+a_1^-(\pi^-a_1^+)\to e^+e^-$ is seen to
dominate over the reaction $\rho^+\rho^-\to e^+e^-$
and is most important in the invariant mass
region 1.5 GeV $< M <$ 3.0 GeV: the possible window for observing the
quark-gluon plasma in
ultrarelativistic nucleus-nucleus collisions.
As expected, the reaction $a_1^+a_1^-\to e^+e^-$
becomes important only at higher invariant masses.
Similar conclusions are reached at
$T=200$ MeV, with an overall enhancement in the dilepton yield.


In summary, we have considered dilepton production from
reactions that involve the $a_1$ meson. We have found
that dilepton production  from the reaction
$\pi^+a_1^-(\pi^-a_1^+)\to e^+e^-$
is more important  than that  from $\rho^+\rho^-$ annihilation.
This observation is independent of the temperature of the hadronic matter.
As in photon production from the hadronic matter,
the $a_1$ meson is important
in dilepton production  from the hot hadronic matter as well.
This fact raises an interesting question about the role
of heavy mesons in hadronic matter. Will we get an enhanced production
of dileptons and photons
as we include other heavier mesons? Are heavier mesons more important than
the light ones in determining the photon and dilepton production rates?
This probably is not the case as
the dilepton production  rates from  reactions that involve
the  $\phi$ meson have been shown in ref. \cite{gale}
to be negligible compared
with the production rate from the reaction $\rho^+\rho^-\to e^+e^-$.
This might indicate that
the enhancement observed in the present study
is just due to the $a_1$ meson.
In this respect,  it will be of interest to study photon production
from strange particles.  Also, it would be useful to extend
 the present study to the case of $SU(3)\times SU(3)$ symmetry.

There are higher order contributions to
dilepton production from
hot hadronic matter \cite{lichard},
which are related to processes similar to those
considered in photon production \cite{song1}.
We expect that a consistent inclusion of the $a_1$
should also add significantly to the contribution from higher order reactions.
Such studies need to be pursued.

C.~Song and C.~M.~Ko were supported by the National Science Foundation under
Grant No.~PHY-9212209 and by the Welch Foundation under Grant
No.~A-1110.C.~Gale
was supported in part by the Natural Sciences and Engineering Research
Council of Canada, and in part by the FCAR Fund of the Qu\'ebec Government.


 
\newpage

%
%

\begin{figure}
\caption{Lowest order diagrams for dilepton emissions from hadronic matter
         which involve axial vector mesons. }
\label{figone}
\end{figure}

\begin{figure}
\caption{Dilepton production  from the hot hadronic matter at $T$= 150 MeV;
         the dotted and dashed curves  are contributions
         from $\pi^+\pi^-\to e^+e^-$ and $\rho^+\rho^-\to e^+e^-$,
         respectively,
         while the solid curves are from
         $\pi^+a_1^-(\pi^-a_1^+)\to e^+e^-$ and $a_1^+a_1^-\to e^+e^-$.}
\label{figtwo}
\end{figure}

\begin{figure}
\caption{Same as the Figure 2 for $T=200$ MeV.}
\label{figthree}
\end{figure}

\end{document}